\documentclass{PoS}
\newcommand\POWHEG{{\tt POWHEG}}
\newcommand\POWHEGBOX{{\tt POWHEG BOX}}

\newcommand\HERWIG{{\tt HERWIG}}

\newcommand\PYTHIA{{\tt PYTHIA}}

\newcommand\ALPGEN{{\tt ALPGEN}}

\newcommand\MCatNLO{{\tt MC@NLO}}

\title{Recent developments in POWHEG}

\ShortTitle{Recent developments in POWHEG}

\author{\speaker{Paolo Nason}%\thanks{A footnote may follow.}
\\
        INFN, Sez. di Milano Bicocca\\
        E-mail: \email{Paolo.Nason@mib.infn.it}}

\abstract{I review recent developments in \POWHEG{}, a method
for interfacing parton-shower generators with NLO QCD
computations. I illustrate recent progress in understanding several
features of the method, and in clarifying similarity
and differences with respect to \MCatNLO{}.
Furthermore, I briefly describe a recently introduced
framework, the \POWHEGBOX{}, that allows the automatic
\POWHEG{} implementation of any given NLO calculation,
and has been recently applied to $Z+{\rm jet}$ production
and to Higgs production via vector--boson fusion.}

\FullConference{RADCOR 2009 - 9th International Symposium on Radiative Corrections (Applications of Quantum Field Theory to Phenomenology) ,\\
		 October 25 - 30 2009\\
		 Ascona, Switzerland}

\begin{document}
The \POWHEG{} method is a prescription for interfacing NLO
calculations with parton shower generators. It was first suggested in
ref.~\cite{Nason:2004rx}, and was described in great detail in
ref.~\cite{Frixione:2007vw}.  Until now, the \POWHEG{} method has been
applied to $Z$ pair hadroproduction~\cite{Nason:2006hfa},
heavy-flavour production~\cite{Frixione:2007nw}, $e^+ e^-$
annihilation into hadrons~\cite{LatundeDada:2006gx} and into top
pairs~\cite{LatundeDada:2008bv}, Drell-Yan vector boson
production~\cite{Alioli:2008gx,Hamilton:2008pd}, $W'$
production~\cite{Papaefstathiou:2009sr}, Higgs boson production via
gluon fusion~\cite{Alioli:2008tz,Hamilton:2009za}, Higgs boson
production associated with a vector boson
(Higgs-strahlung)~\cite{Hamilton:2009za}, single-top
production~\cite{Alioli:2009je} $Z+1$~jet
production~\cite{POWHEG_Zjet}, and, very recently,
Higgs production in vector boson fusion~\cite{Nason:2009ai}.
Unlike \MCatNLO{}
\cite{Frixione:2002ik}, \POWHEG{} produces events with
positive (constant) weight, and, furthermore, does not depend on the
subsequent shower Monte Carlo program. It can be easily interfaced to
any modern shower generators and, in fact, it has been interfaced to
\HERWIG{}~\cite{Corcella:2000bw,Corcella:2002jc} and
\PYTHIA{}~\cite{Sjostrand:2006za} in
refs.~\cite{Nason:2006hfa,Frixione:2007nw,Alioli:2008gx,Alioli:2008tz,
  Alioli:2009je}.

\section{The concept}
The basic concept of \POWHEG{}
is better clarified by considering a process with a single massless
coloured parton involved (one can consider, for example,
semileptonic top decay, neglecting the $b$ mass).
In a shower Monte Carlo, the radiation of a final state
light parton is generated with an algorithm that resums all leading log corrections to
the Born process. The hardest emission in a shower Monte Carlo is generated
according to the
formula~\cite{Nason:2004rx}
\begin{equation}\label{eq:mchardest}
d \sigma = B(\Phi_B) d\Phi_B
\left[\Delta^{\rm MC}_{t_0}(\Phi_B)+\Delta^{\rm MC}_t(\Phi_B)
\frac{R^{\rm MC}(\Phi)}{B(\Phi_B)}
d\Phi_r^{\rm MC}\right],
\end{equation}
where $t$ is the radiation transverse momentum, $t_0$ is the minimum
allowed value for $t$ (typically of the order of a hadronic scale),
$B(\Phi_B)d\Phi_B$ is the
Born differential cross section, and $R^{\rm MC} d\Phi_B\Phi_r$ is the real
radiation differential cross section in the Monte Carlo (MC from now on)
approximation. 
It is assumed that the full phase space
including radiation is parametrized in term of the Born phase space $\Phi_B$
and the radiation phase space $\Phi_r$, i.e. $\Phi=\Phi(\Phi_B,\Phi_r)$.
In typical MC's,
the radiation phase space is determined by three variables characterizing
the collinear splitting process, like, for example, the splitting angle,
the momentum fraction and the azimuth. The radiation transverse momentum
$t$ is a function of $\Phi_B$ and $\Phi_r$. It can be defined as the momentum
component of the radiated parton orthogonal to the momentum of the radiating
parton.
The MC Sudakov form factor
\begin{equation}\label{eq:mcsuda}
\Delta^{\rm MC}_{t_l}(\Phi_B)=\exp\left[-\int_{t>t_l} \frac{R^{\rm MC}(\Phi)}{B(\Phi_B)} d\Phi_r^{\rm MC}
\right]
\end{equation}
represents the probability for not having radiation harder than $t_l$.

The basic idea in \POWHEG{} is to improve formula (\ref{eq:mchardest}) in such
a way that NLO accuracy is reached. One replaces formula (\ref{eq:mchardest})
with the following one
\begin{equation}\label{eq:pwghardest}
d\sigma =\bar{B}(\Phi_B)d\Phi_B\left[\Delta^s_{t_0}(\Phi_B)+\Delta^s_t(\Phi_B)
\frac{R^s(\Phi)}{B(\Phi_B)} d\Phi_r^{\rm MC}\right]+[R(\Phi)-R^s(\Phi)]d\Phi,
\end{equation}
where $\Phi$ is the full phase space, with $d\Phi=d\Phi_B d\Phi_r$,
and $R$ is the exact radiation cross section.
We require that $R\rightarrow R^s$ in the soft and collinear limit,
and that $R^s\le R$, so that the last contribution in the square bracket
of eq.~(\ref{eq:pwghardest}) is non-negative.
The choice $R^s=R$ is also possible, and it is quite common.
We have also defined
\begin{equation}\label{eq:bbar}
\bar{B}(\Phi_B)=B(\Phi_B)+\left[V(\Phi_B)+\int R^s(\Phi) d\Phi_r\right],
\end{equation}
where $V$ is the virtual NLO correction to the Born process.
Notice that soft and collinear singularities in $V$ cancel against those
arising from the integral of $R^s$ in the square bracket of eq.~(\ref{eq:bbar}).
The Sudakov form factor is now
\begin{equation}
\Delta^{\rm s}_{t_l}(\Phi_B)=\exp\left[-\int_{t>t_l} \frac{R^{\rm s}(\Phi)}{B(\Phi_B)} d\Phi_r\;.
\right]
\end{equation}
The phase space factorization in the \POWHEG{} formula needs not to match that
of any shower Monte Carlo. One only requires that in the soft and collinear
limit the full phase space $\Phi$ is related to the Born phase space $\Phi_B$
in the correct way, i.e. they are identical in the soft limit once the soft
particle is removed, and they are identical in the collinear limit once the
collinear particles are merged. The \POWHEG{} formula (\ref{eq:pwghardest}) can
be viewed as an improvement of the Monte Carlo formula (\ref{eq:mchardest}),
such that the Born cross section is replaced with an NLO inclusive cross section, and high transverse momentum radiation is corrected so that it becomes
exact at large angles. In fact, for large $t$ the Sudakov form
factor becomes 1, and the \POWHEG{} cross section reduces to
\begin{equation}\label{eq:pwghardestlarget}
d \sigma=\bar{B}\times \frac{R^s}{B} d\Phi+[R-R^s]d\Phi\approx R d\Phi,
\end{equation}
since $\bar{B}/B= 1+{\cal O}(\alpha_s)$. At small $t$ the \POWHEG{} formula
becomes equal to that of a standard shower MC, up to higher order
terms. However, since by construction
\begin{equation}
\Delta^{\rm s}_{t_0}+\int \theta(t-t_0)\;\Delta^{\rm s}_t
\frac{R^s(\Phi)}{B(\Phi_B)} d\Phi_r^{\rm MC}=1
\end{equation}
the \POWHEG{} formula maintains NLO accuracy for integrated (i.e. inclusive)
quantities.

Notice that the same formula (\ref{eq:pwghardest}) also describes the radiation
of the hardest parton in \MCatNLO{}, provided $R^s$ is identified with the
shower Monte Carlo (i.e. \HERWIG{}'s) approximation of the real emission cross
section. In fact, in \MCatNLO{} two types of events are generated, called
${\cal S}$ and ${\cal H}$ events. ${\cal S}$ events correspond to the term
proportional to $\bar{B}$. In \MCatNLO{} the corresponding underlying Born
kinematics is generated with a probability $\bar{B}(\Phi_B) d\Phi_B$, while
the hardest radiation kinematics is generated by the \HERWIG{} shower algorithm.
It was demonstrated in ref.~\cite{Nason:2004rx} that the hardest radiation
in \HERWIG{} corresponds to the factor in square bracket multiplying $\bar{B}$
in eq.~(\ref{eq:pwghardest}). The ${\cal H}$ events correspond instead to the
$R-R^s$ term in eq.~(\ref{eq:pwghardest}). However, since $R^s$ is now given by
the \HERWIG{} shower algorithm, there is no guarantee that the difference
$R-R^s$ should be positive, and this is why negative weighted events are an
essential feature of \MCatNLO{}.

\begin{figure}
\includegraphics[width=0.485\textwidth]{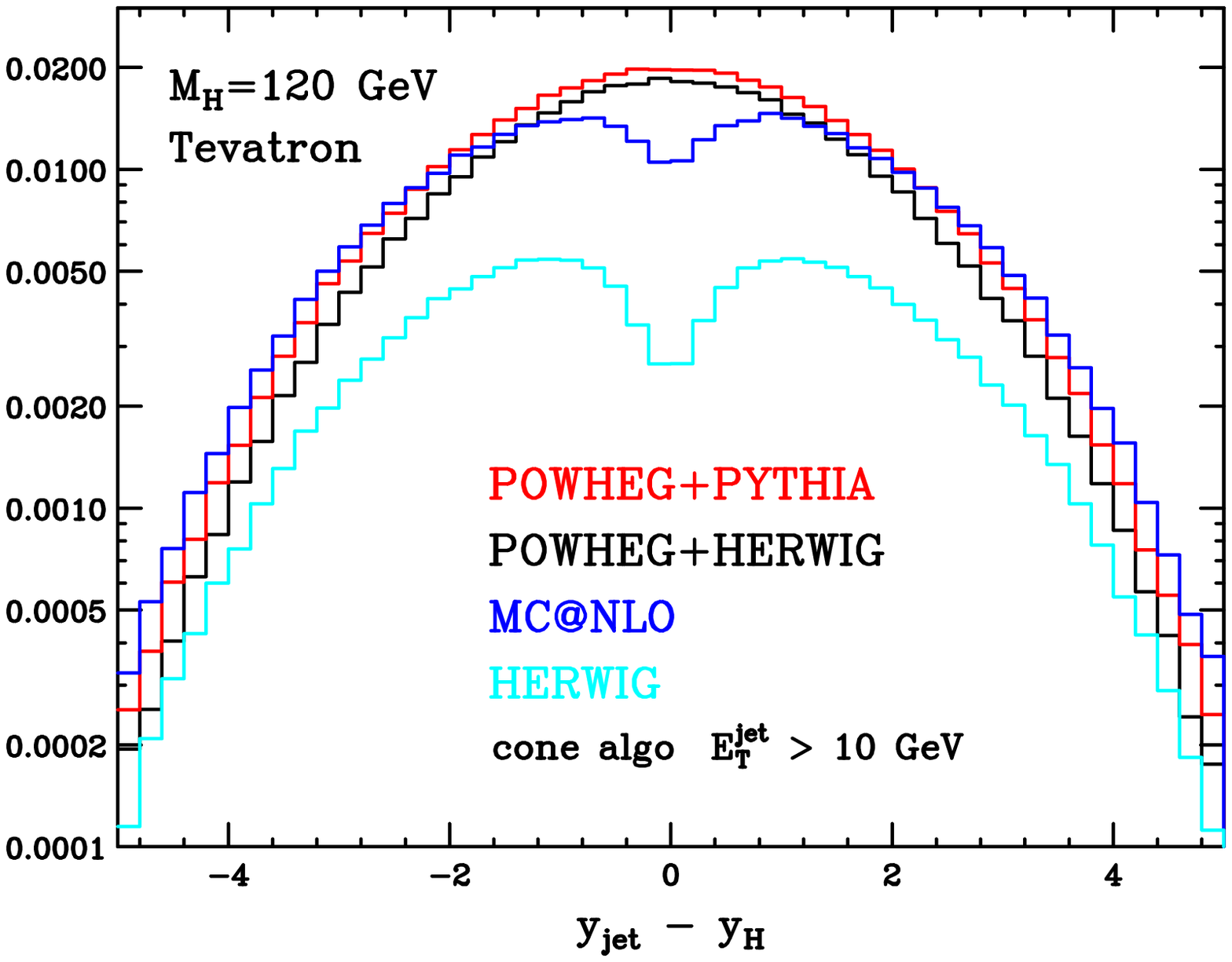}
\includegraphics[width=0.515\textwidth]{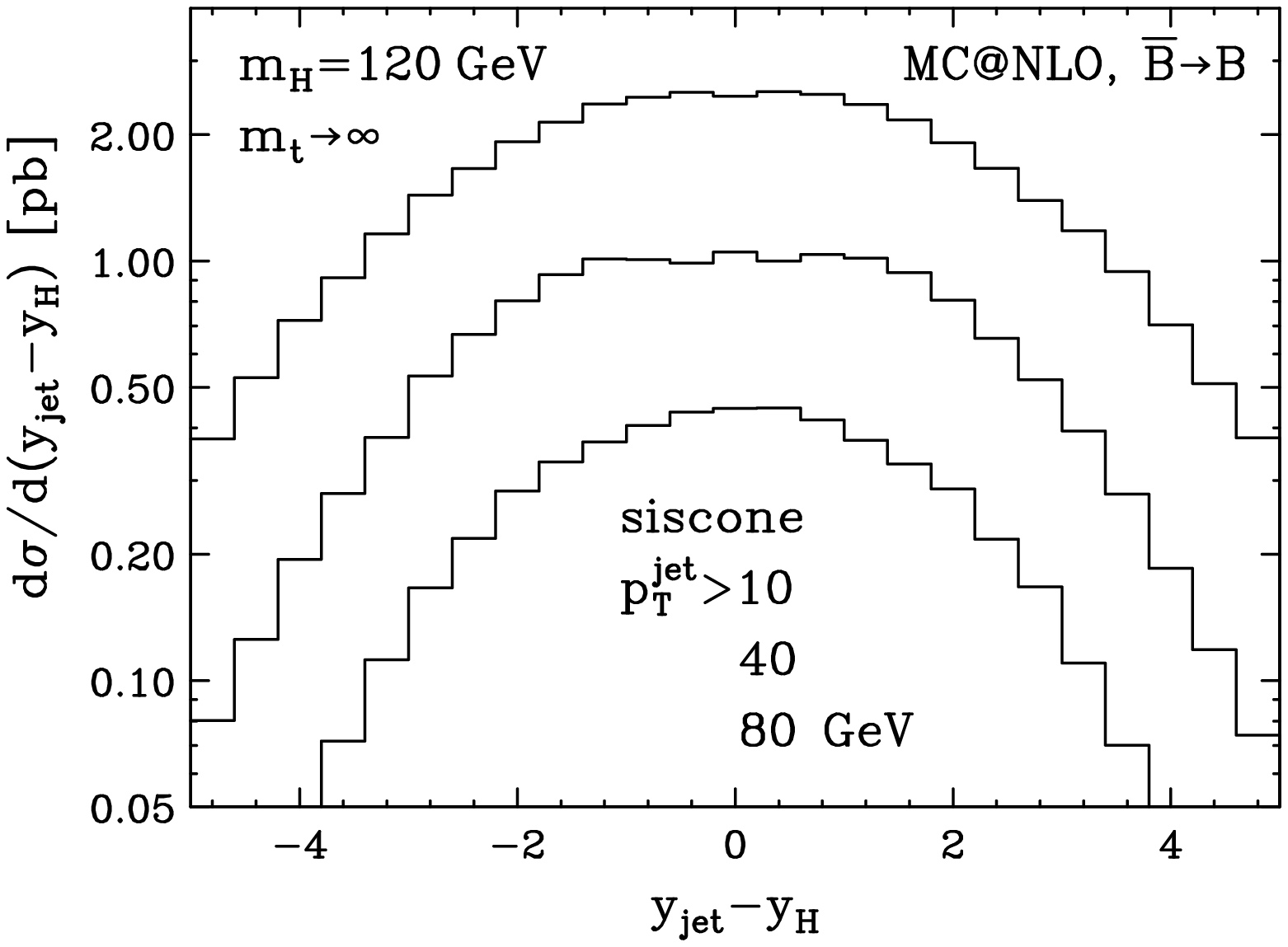}
\caption{Left: rapidity distribution of the hardest jet relative to the
Higgs rapidity. The highest curve at $y=0$ is \POWHEG{}$+$\PYTHIA{}, next is
\POWHEG{}$+$\HERWIG{}, then \MCatNLO{} and then \HERWIG{} alone.
Right: \MCatNLO{}, with $\bar{B}$ replaced by $B$ in the generation
of ${\cal S}$ events.}
\label{fig:higgsjetrap}
\end{figure}

The fact that both in \MCatNLO{} and \POWHEG{} the hardest radiation
can be described by a similar formula has allowed a better understanding
of the
agreement and discrepancies between the two approaches.
First of all, one understands why most distributions compare very well
in the two schemes (see for example \cite{Nason:2006hfa,Frixione:2007nw,%
Alioli:2008gx,Alioli:2009je}), since they are both described
by a similar formula.
A first area of discrepancy has emerged following the work of
ref.~\cite{Mangano:2006rw}. In $t\bar{t}$ production, a dip in the rapidity
distribution of the hardest jet of \MCatNLO{}
was found, that is not present neither in \ALPGEN{} nor in \POWHEG{}.
It was shown later \cite{Alioli:2008gx,Alioli:2008tz} that this
dip is a feature of \MCatNLO{} that is present in several processes.
It is particularly visible in Higgs production, as one can see from
fig.~\ref{fig:higgsjetrap}.
The pure \HERWIG{} result exhibits a dip at zero rapidity.
\MCatNLO{} has a similar dip, although relatively less severe.
No dip is present
in \POWHEG{}. The origin of the dip is easily understood
as a consequence of the presence of the dip in pure \HERWIG{} and of
formula (\ref{eq:pwghardest}). In fact, for large transverse momenta,
formula (\ref{eq:pwghardest}) becomes equal to
\begin{equation}\label{eq:pwghardessymp}
d \sigma=\bar{B}\times \frac{R^s}{B} d\Phi+[R-R^s]d\Phi=
R\, d\Phi+\left[\frac{\bar{B}}{B}-1\right]R^s d\Phi.
\end{equation}
The second term is small in a perturbative sense, being of order $\alpha_s$.
However, in the case of Higgs production, where NLO corrections amount
to 100\%{}, it is in fact of order 1. Since $R^s$ correspond to the \HERWIG{}
approximation to the real cross section, it has a dip that propagates into
the \MCatNLO{} cross section. We have checked this
explicitly. If this argument is correct, by replacing the $\bar{B}$ cross
section with $B$ in \MCatNLO{}, the dip should go away. This is in fact shown
in fig.~\ref{fig:higgsjetrap}, on the right plot.

\begin{figure}
\includegraphics[width=0.515\textwidth]{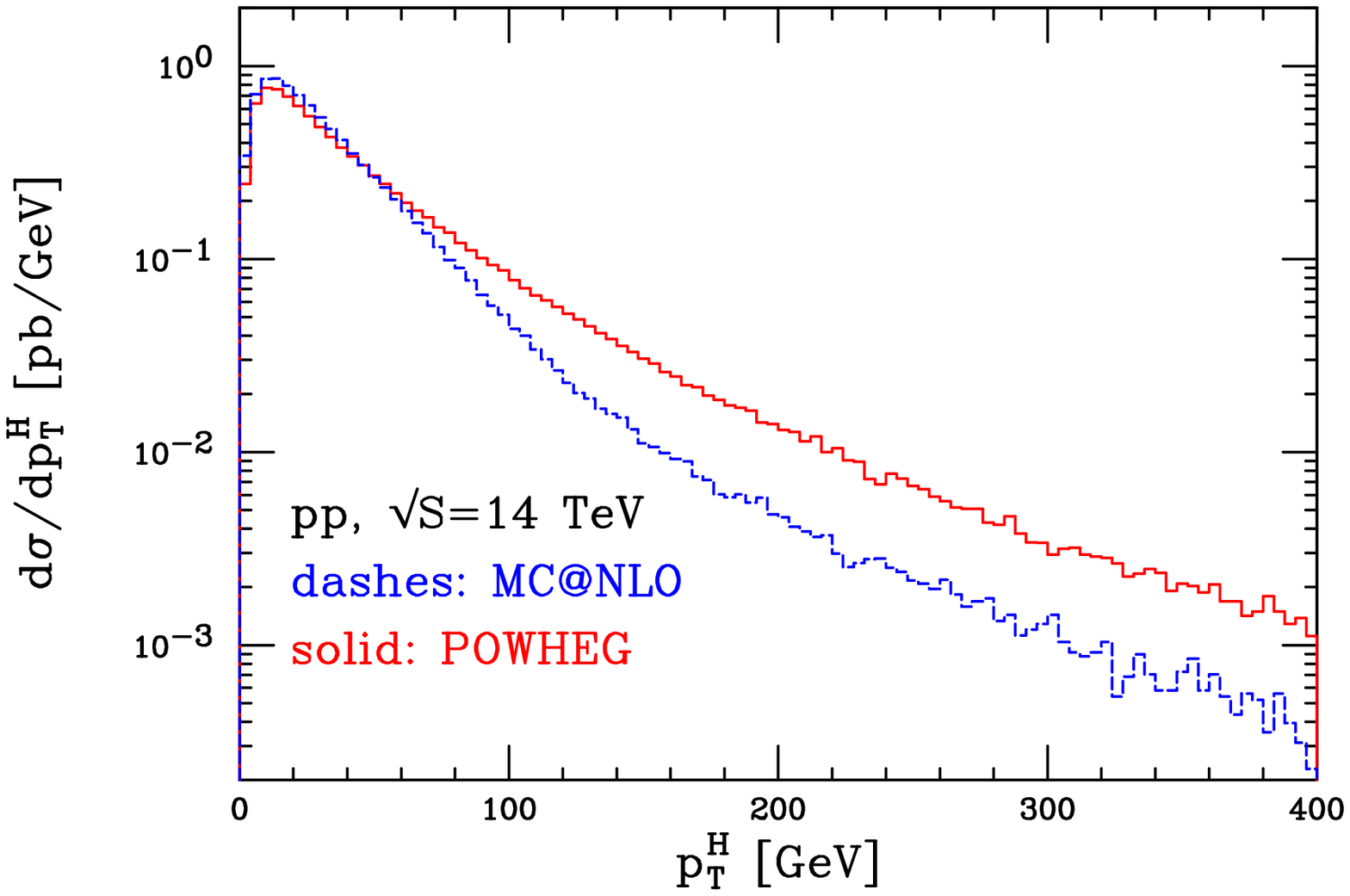}
\includegraphics[width=0.485\textwidth]{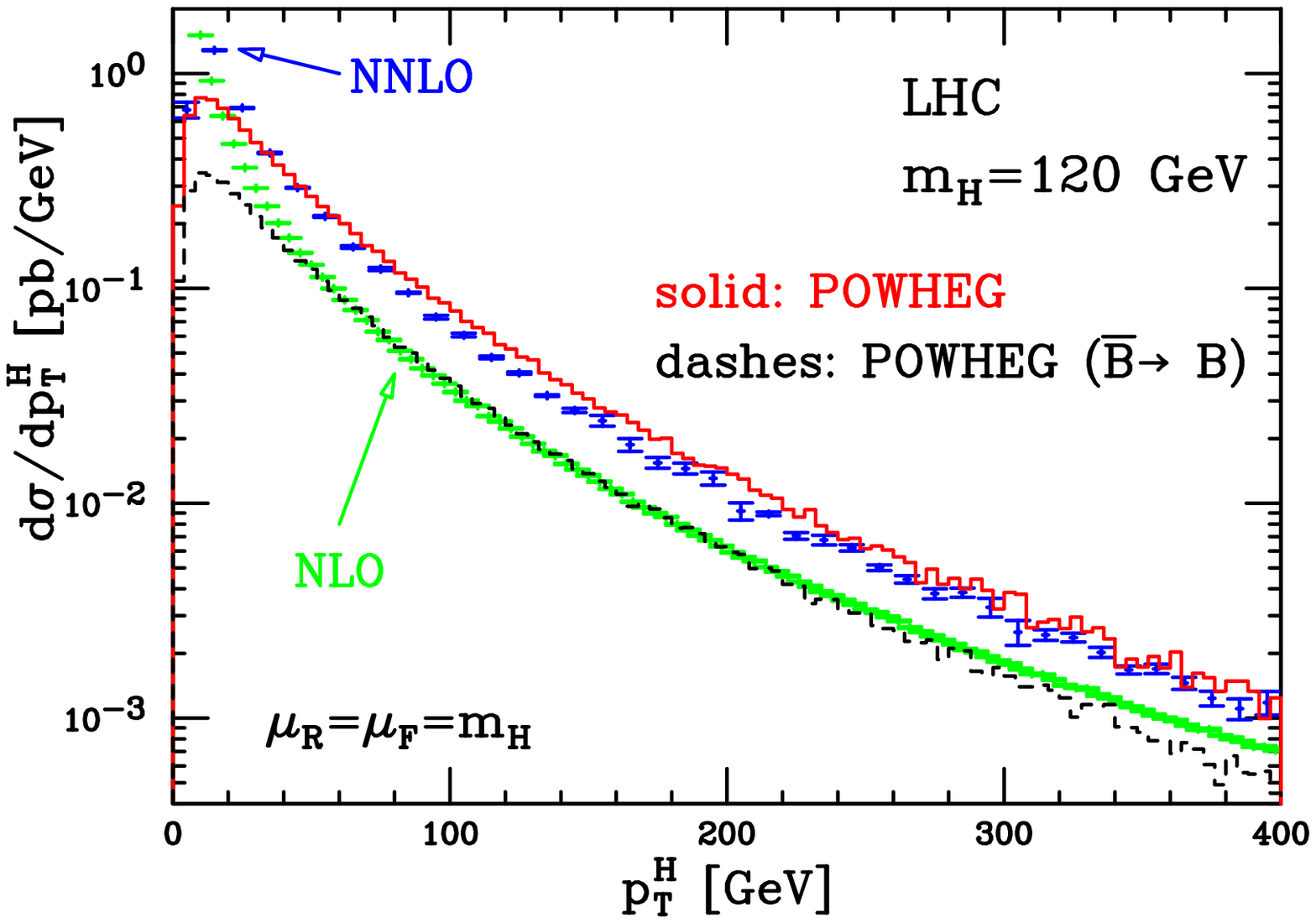}
\caption{Left: rapidity distribution of the hardest jet relative to the
Higgs rapidity. The highest curve at $y=0$ is \POWHEG{}$+$\PYTHIA{}, next is
\POWHEG{}$+$\HERWIG{}, then \MCatNLO{} and then \HERWIG{} alone.
Right: \MCatNLO{}, with $\bar{B}$ replaced by $B$ in the generation
of ${\cal S}$ events.}
\label{fig:pthiggs}
\end{figure}

Another important discrepancy is found in the Higgs $p_T$ distribution
in the $gg\to h$ process, displayed in the left plot of
fig.~\ref{fig:pthiggs}.
At high $p_T$ the \POWHEG{} and \MCatNLO{}
spectra depart sensibly. The origin of this difference can again be understood
starting from formula (\ref{eq:pwghardessymp}). In the \POWHEG{} implementation
of Higgs production, the choice $R^s=R$ was made, so that formula (\ref{eq:pwghardessymp}) yields
\begin{equation}\label{eq:pwghardessymp1}
d \sigma=\frac{\bar{B}}{B}\times R\, d\Phi\;.
\end{equation}
Because of the large Higgs production NLO corrections, the $\bar{B}/B$ factor,
instead of being near 1 is close to 2. Thus NLO corrections amplify the whole
$p_T$ spectrum in \POWHEG{}. Of course, we could have chosen a different $R^s$
in order to reduce this effect, but in view of the plot on the right
in fig.~\ref{fig:pthiggs}, this choice was maintained, since the $p_T$
spectrum produced in this way is closer to the NNLO result.
In \MCatNLO{} this amplification works only for the ${\cal S}$ events,
that become negligible at large $p_T$.
Notice that in fig.~\ref{fig:pthiggs} a mild kink is visible at 80-110~GeV in
the \MCatNLO{} curve,
signalling the separation of the ${\cal S}$ and ${\cal H}$ dominated regimes.
\section{The \POWHEGBOX{}}
In ref.~\cite{Frixione:2007vw}, the general formulation of \POWHEG{} for
complex processes was given, dealing with all complications having to do
with the presence of several massless coloured lines, both in the final
and in the initial state. In this case formula~(\ref{eq:pwghardest}) should
be applied to each radiating line.
What a regular shower Monte Carlo does, starting from a Born process, is to
let each coloured massless line radiate more light partons.
Within \POWHEG{} one does something similar. However, since in this case
the radiation cross section is the exact cross section, rather than
an Altarelli-Parisi approximation to it, we work the other way around,
decomposing the full real cross section into a sum of terms, each of them
singular in only one collinear region. We then adopt a different
phase space parametrization $d\Phi\to d\Phi_Bd\Phi_r$ for each singular
region. After this is done, radiation is generated from each collinear
region, but only the hardest radiation is retained, and the event is generated
according to its kinematics.
This procedure reaches a considerable level of complexity, so we decided
to implement it in a general way, in order to ease the \POWHEG{} implementation
of complex processes. This project, named the \POWHEGBOX{} has been
completed~\cite{ANOR2010-1}, and will soon been published. It has been used to
reproduce the previous implementation of $W/Z$ and single top production.
Furthermore, it has been used to implement $Z+{\rm jet}$
production~\cite{POWHEG_Zjet},
and Higgs production in vector boson fusion~\cite{Nason:2009ai}.
This last two processes are the first ones not already
present in \MCatNLO{}.

Internally, the \POWHEGBOX{} implements the NLO corrections using the
FKS subtraction framework \cite{Frixione:1997np,Frederix}.
The author of an NLO calculation, willing to implement
it into the \POWHEGBOX{} framework, needs only to supply the following
ingredients:
a) the Born phase space; b)
the list of Born and Real processes in a specified format;
c) the Born squared amplitudes $B=|{\cal M}^2|$, the colour correlated one
 $B_{ij}$, and the spin correlated one $B_{j,\mu_j,\mu'_j}$, for each Born
subprocess (these ingredient are typically needed in NLO calculations
performed with the subtraction method, and, for example, are generated
automatically in the MadDipole package \cite{Frederix:2008hu});
d) the real squared amplitude, for all relevant partonic processes
 (this is easily obtained from programs like MadGraph \cite{Alwall:2007st});
e) the finite part of the virtual amplitude for all relevant partonic processes,
which is the only ingredient that may require dedicated work.
The \POWHEGBOX{} does all the rest, including the automatic implementation
of the FKS subtractions.
\section{Conclusions}
Much progress has been achieved in \POWHEG{} in this past year.
Several features of the method have been better understood in simple
terms, and its comparison with \MCatNLO{} has been considerably clarified.
An important step towards automation has been achieved with the
construction of the \POWHEGBOX{}. Thanks to this framework, a relatively
complex process like VBF Higgs production has been implemented in a very short
time. It is our hope that the \POWHEGBOX{} will allow authors of NLO
calculations to implement their work in the \POWHEG{} framework with
a limited effort.
\bigskip\\
{\bf Acknowledgements}\\
All results presented in this talk have been obtained in collaboration
with S. Alioli, C. Oleari and E. Re.

\end{document}